\documentclass[11pt]{article}
\usepackage[margin=1in]{geometry}
\usepackage{amsmath,amssymb}
\usepackage{graphicx}
\usepackage{xcolor}
\usepackage[labelsep=period]{caption}
\usepackage{url}
\usepackage[numbers,sort&compress]{natbib}
\usepackage{lineno}
\usepackage[hidelinks]{hyperref}

\graphicspath{{figs/}}

\title{Multistable elastic pixels: nematic elasticity for non-volatile positioning of colloidal 
optical scatterers}
\author{Jed-Joan S. Edziah$^{1}$, Brian Edwards$^{2}$, Nader Engheta$^{2,*}$, Kathleen J. Stebe$^{1,*}$\\[4pt]
\small $^1$Chemical and Biomolecular Engineering, University of Pennsylvania,\\ \small 220 South 33rd Street, Philadelphia, PA 19104, USA\\
\small $^2$Electrical and Systems Engineering, University of Pennsylvania,\\ \small 200 South 33rd Street, Philadelphia, PA 19104, USA\\
\small $^*$engheta@seas.upenn.edu; kstebe@seas.upenn.edu}

\begin{document}
\maketitle

\begin{abstract}
Non-volatile reconfiguration of optical devices, in which a configuration persists without continuous energy input, remains a challenge. We introduce multistable elastic pixels (MEPs), lithographically defined pores filled with a nematic liquid crystal, whose curved ends capture a scattering inclusion in one of two mechanically stable sites. Simulations and experiments show that pore geometry and homeotropic surface anchoring define the energy minima and the height of the barrier between them. This barrier far exceeds thermal energies, stabilizing a scatterer location absent an external field. Colloidal inclusions switch reversibly between sites under transient electric fields. Using arrays of static scatterers, we show that the two positional states produce distinct responses at visible wavelengths, demonstrating the potential of MEPs for reconfigurable devices and metasurfaces. Unlike phase-change and liquid-crystal-tuned approaches, which switch a material surrounding fixed scatterers, MEPs use the nematoelastic energy landscape to move the optical element itself. 
\end{abstract}


\section*{Introduction}

Many advanced optical technologies rely on control over an optical element's position. In electronic paper, electric fields translate charged pigment particles to set pixel color, and the image persists without power \cite{comiskey1998}. Optical security tags encode information in the arrangement of scatterers \cite{geng2016}; such tags that can be rewritten afford additional security \cite{kim2026,he2025}. Metasurfaces, planar arrays of engineered scatterers designed to shape the amplitude, phase and polarization of incident electromagnetic (EM) waves \cite{engheta2011,yu2014}, are another prominent example. Metasurfaces have been realized as lenses, beam splitters, beam steering devices and polarization optics \cite{arbabi2023,li2024pol}, and advances in lithography now support large-area and commercial fabrication \cite{huang2022}. Most metasurfaces are static, with an optical function fixed at fabrication. Reconfigurable metasurfaces, in which the scatterers, their positions or their surroundings can be altered after fabrication, would allow a single device to access multiple optical functions \cite{shaltout2019}. Non-volatility, in which a configuration persists once written without continuous energy input, is valuable wherever a configuration must be held for long periods \cite{gu2023}.

Reconfigurability has been imparted by mechanical, electrical and thermal actuation. Stretching elastomeric substrates tunes inclusion spacing and focal length \cite{kamali2016}, and microelectromechanical systems (MEMS) translate or displace metasurface elements under applied bias \cite{he2019mems,han2020,ou2013,kwon2022}. Electrostatic gating of conducting oxides and graphene modulates carrier density and thereby the amplitude and phase of scattered light \cite{huang2016,sherrott2017,squires2022}. These approaches can be fast, but the altered state persists only while the strain or bias is maintained. At optical frequencies, non-volatile reconfiguration has been achieved almost exclusively with chalcogenide phase-change materials, whose amorphous and crystalline phases are both metastable at room temperature \cite{wangQ2016,zhangY2021,abdollah2022}. Phase-change metasurfaces now support individually addressable pixels \cite{fang2024,popescu2026}. However, switching requires melt-quench or annealing pulses delivered by integrated heaters, and the optical function remains tied to the refractive index of the switching material \cite{fang2024}. Mechanical latching provides an alternative form of memory and has been demonstrated with bistable origami unit cells at microwave frequencies \cite{zhangD2026}, but it is difficult to realize at the micrometer scale.

Liquid crystals (LCs) are widely used in reconfigurable metasurfaces. Typically, the birefringence of a nematic liquid crystal (NLC) surrounding fixed resonators is reoriented by an applied field, tuning their resonances and hence the amplitude, phase or polarization of scattered light \cite{buchnev2013,decker2013,komar2017,sharma2022}. This strategy underlies metasurface spatial light modulators with micrometer-scale pixels \cite{li2019,moitra2023} and pixel-addressable holographic displays \cite{liJ2020}. In these devices the LC serves as a tunable optical medium; the director relaxes once the field is removed, and the optical state is volatile. LCs have also been used to organize optically active nanoparticles. Metallic nanorods align with the director and can be reoriented by electric or optical fields \cite{zhang2015,sang2023}, and topological defects template the assembly of nanoparticles into rings and chains \cite{lee2016,tasinkevych2023,coursault2012}, with tunable plasmonic responses \cite{lee2016,coursault2012}. Reconfiguration in these systems again relies on sustained fields or thermal actuation of the host, and self-assembled structures lack registry with a lithographically defined lattice.

Nematic elasticity also provides a means of storing a configuration mechanically. In bistable nematic displays, surface gratings or microposts with prescribed anchoring stabilize topologically distinct director configurations that persist without power and are switched by transient fields \cite{zbd2022,kitson2002,majumdar2007}. Confinement in wells and porous media similarly supports multiple metastable director states \cite{tsakonas2007,araki2011,serra2013}. In these systems, however, the stored state is the director field itself, and optical contrast arises from its birefringence. Boundaries also act on inclusions dispersed in a nematic. In the nematic lock-and-key mechanism, colloids find docking sites near curved boundaries with similar anchoring, minimizing the distortion in the domain \cite{silvestre2004,luo2016pre,luo2016,luo2018,boniello2019}. Such geometry-guided positioning is driven solely by the minimization of elastic energy and requires no continuous energy input once an inclusion is docked.

Here we combine these elements in multistable elastic pixels (MEPs), NLC-filled pores whose geometry and anchoring define two mechanically stable docking sites for a scattering inclusion, separated by an elastic energy barrier. An electric field translates the inclusion from one site to the other, and the inclusion remains in place once the field is removed. Unlike phase-change and LC-tuned metasurfaces, in which the scatterers are fixed and a material property near them is switched, in MEPs the scatterers themselves move.  The NLC provides the elastic energy landscape, so the composition of the scatterer is a design parameter. We design bistable pillbox and peanut-shaped MEPs and use Landau--de Gennes simulations to show that pore geometry and homeotropic anchoring set reconfiguration barriers that far exceed thermal energies. Experiments with colloidal inclusions confirm bistability, reversible switching under transient fields, and the geometric trends predicted by simulation. Finally, using lithographically patterned arrays with scatterers placed at one site or the other, we show that the two positional states give distinct optical responses at visible wavelengths.

\begin{figure}[t]
\centering
\includegraphics[width=\textwidth]{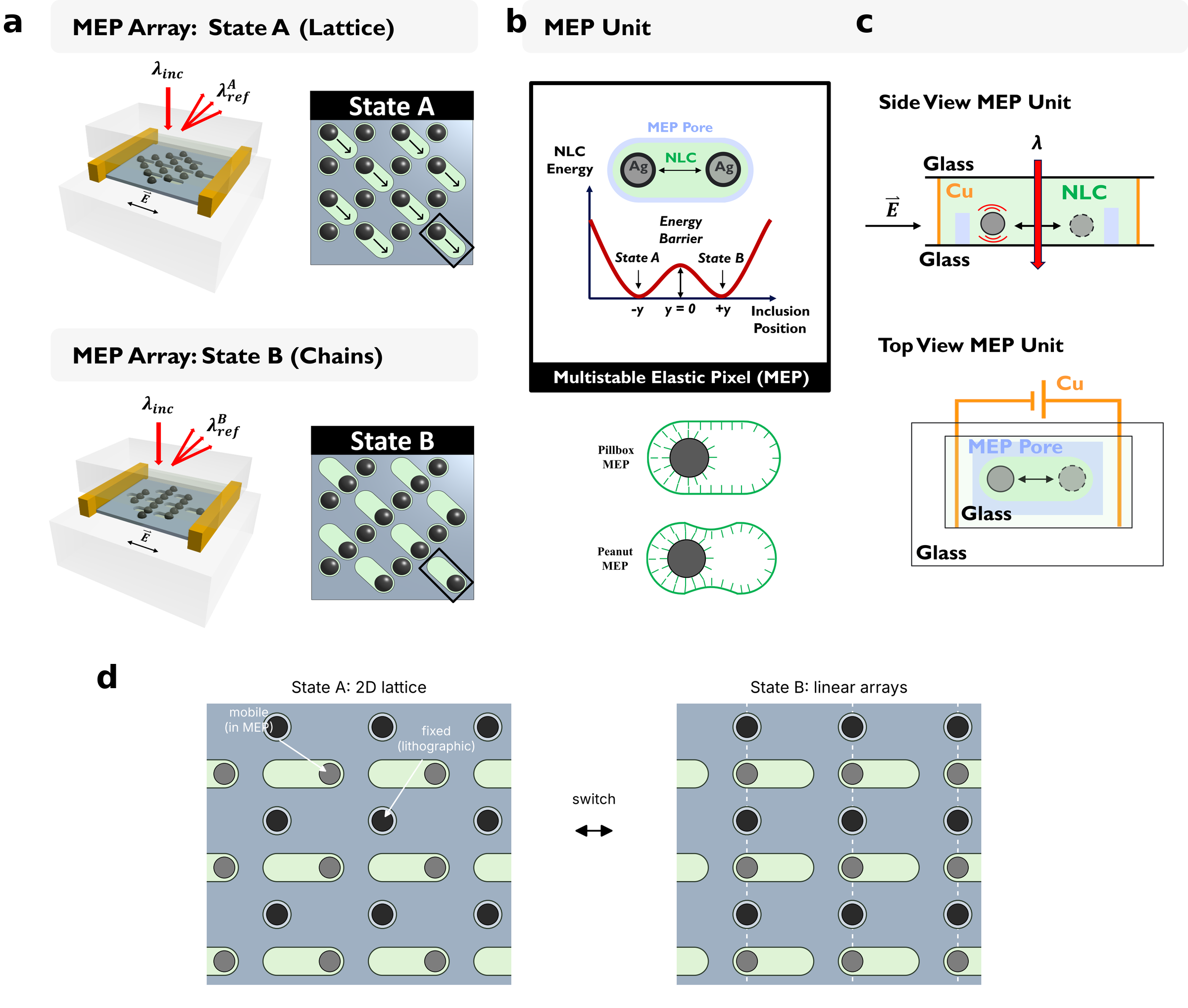}
\caption{\textbf{Multistable elastic pixels (MEPs).} \textbf{a} MEP arrays in State A (lattice) and State B (chains), with incident and reflected light and the in-plane switching field $\vec{E}$. \textbf{b} The MEP unit. An inclusion in an NLC-filled pore has two docking sites (States A and B) separated by an elastic energy barrier; pillbox and peanut pore geometries. \textbf{c} Side and top views of an MEP unit with Cu electrodes. \textbf{d} Fixed and mobile scatterers. Scatterers placed lithographically in circular pores remain fixed; scatterers in MEPs are mobile. Switching the mobile scatterers converts a two-dimensional lattice (State A) into linear arrays (State B).}
\label{fig:concept}
\end{figure}

Figure~\ref{fig:concept} illustrates the MEP platform. An NLC cell consists of a nematic layer confined between substrates that impose prescribed surface anchoring, with electrodes that apply an external switching field. We introduce bistable MEPs in two classes, pillbox and peanut, in which curved ends are connected either by straight sidewalls or by a constricted waist; the two docking sites define State A and State B. In schematic and simulated representations, dashed lines depict the orientation of the nematogens, which are perpendicular to all bounding surfaces. Arranged on a surface, MEPs define diffractive arrays whose scattering depends on which docking site each inclusion occupies. Because scatterers can also be placed lithographically at fixed positions, MEP arrays could combine fixed and mobile scatterers in a single device. Figure~\ref{fig:concept}a,d illustrates a proposed two-state device of this kind, in which fixed scatterers occupy circular pores, mobile scatterers occupy MEPs, and switching the mobile scatterers converts a two-dimensional lattice (State A) into linear arrays (State B).

\section*{Results}

\subsection*{Multistable reconfiguration mechanism}

The reconfigurable behavior of the MEP platform is governed by elastic distortions in the surrounding nematic liquid crystal, which generate forces that make each MEP bistable. The orientation of the rod-like nematogens is described by a headless vector $\mathbf{n}$, the director \cite{degennes1995}, and distortions of the director from uniform alignment cost elastic energy. Surface anchoring sets the preferred director orientation at the inclusion and the MEP walls, so the elastic energy depends on the position of the inclusion within the pore. An inclusion displaced from an equilibrium site increases the distortion and elastic cost; the resulting restoring force drives it back to the nearest energy minimum.

\subsection*{Simulated MEPs}
We model director fields and elastic free energies in the MEP geometry using a continuum framework (see Methods) \cite{sussman2019}. The total elastic free energy $F$ is the sum of the bulk free energy cost of distortions of the nematic director field $\mathbf{n}$ and the surface free energy at the walls of the pore and the surface of the colloidal inclusion,
\begin{equation}
\label{eq:total_free_energy}
F=\int_{V} f_b\,\mathrm{d}V+\int_{S} f_s\,\mathrm{d}S .
\end{equation}
All surfaces impose strong homeotropic anchoring, with surface energy density
\begin{equation}
\label{eq:surface_free_energy}
f_s=\frac{W}{2}\left[1-\left(\mathbf{n}\cdot\boldsymbol{\nu}\right)^2\right],
\end{equation}
where $\boldsymbol{\nu}$ is the surface normal and the anchoring strength $W$ is large.

The costs to splay, twist and bend  in the director field are characterized by elastic constants $K_1$, $K_2$ and $K_3$,  respectively. 
In the one-constant (1C) approximation, which assumes that these costs are comparable, they are replaced by a single constant $K$.  
To resolve the three-dimensional nematic configurations within MEPs, simulations are performed within a Landau--de Gennes (LdG) Q-tensor framework, which captures spatial variations in the director and resolves the scalar order parameter. 
For uniaxial nematics,
\begin{equation}
\mathbf{Q} = \frac{3}{2} S \left(\mathbf{n}\mathbf{n} - \frac{1}{3}\mathbf{I}\right),
\end{equation}
where $S$ is the scalar order parameter and $\mathbf{I}$ is the identity tensor. The free energy expressed in terms of $\mathbf{Q}$ is minimized with respect to both the director and the magnitude of order \cite{ravnik2009}. Topological defects appear as regions of reduced order, and stable inclusion positions appear as local minima of the total free energy \cite{kralj2021}. The LdG elastic constants in our simulations are set from the one-constant value of $K$ for 4-cyano-4$'$-pentylbiphenyl (5CB), the NLC used throughout this work. Results with the three elastic constants of 5CB are given in  Supplementary Note 1. 

Within the MEP, competition between bulk elastic distortions and surface anchoring defines an energy landscape for the inclusion. Local minima in the landscape correspond to stable docking sites, and energy barriers separate them. As in lock-and-key docking \cite{luo2016,luo2016pre,luo2018}, inclusions move toward curved boundaries with similar anchoring to minimize distortion in the domain.

\begin{figure}[p]
\centering
\includegraphics[width=\textwidth]{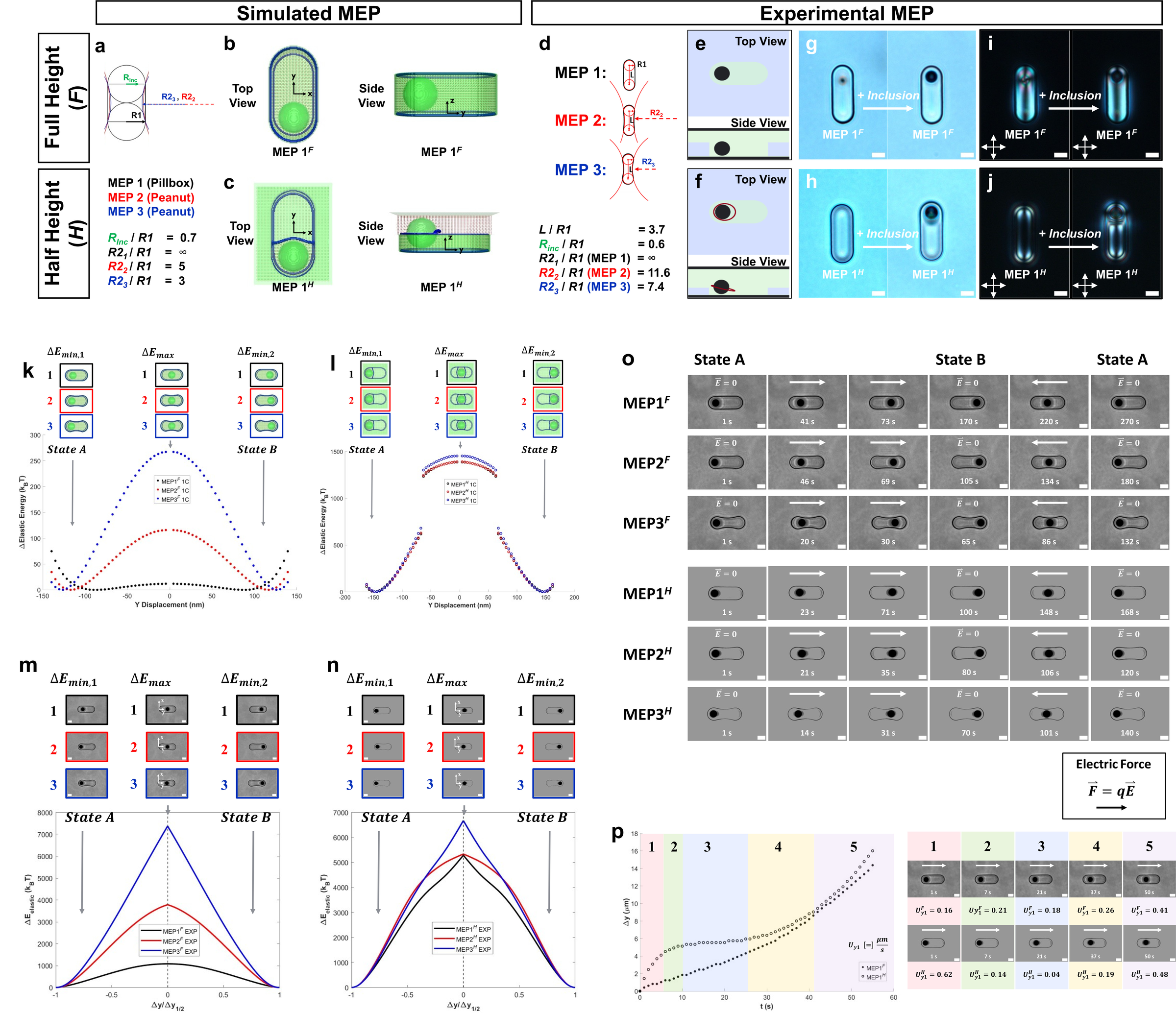}
\caption{\textbf{Simulated and experimental MEPs.} \textbf{a} Simulated pore geometry and parameters for full height (F) and half height (H) MEPs. \textbf{b,c} Simulated director fields, top and side views, for a spherical inclusion of radius 94.5~nm docked in MEP1$^F$ (\textbf{b}) and MEP1$^H$ (\textbf{c}); all surfaces impose homeotropic anchoring. In simulated director fields, green objects are boundaries (inclusion and pore surfaces), blue objects are defects and red dashed lines are nematogens. \textbf{d} Experimental MEP geometries and parameters. \textbf{e,f} Schematics of docked inclusions in full (\textbf{e}) and half (\textbf{f}) height pores. \textbf{g--j} Bright-field (\textbf{g,h}) and crossed-polarizer (\textbf{i,j}) images of empty pores and pores with docked Ag-coated silica spheres of radius $\sim$4.5~\textmu m. \textbf{k,l} Simulated elastic energy versus inclusion displacement, full (\textbf{k}) and half (\textbf{l}) height. \textbf{m,n} Experimental elastic energy profiles inferred from relaxation, full (\textbf{m}) and half (\textbf{n}) height. \textbf{o} Reversible switching, State A $\to$ B $\to$ A, for all six designs under a transient DC field. \textbf{p} Inclusion displacement $\Delta y$ versus time for MEP1$^F$ (filled) and MEP1$^H$ (open) during field-driven reconfiguration; shaded regions 1 to 5 are tabulated at right with representative micrographs and the average velocity $U_{y1}$ (\textmu m\,s$^{-1}$) in each region. Scale bars, 10~\textmu m.}
\label{fig:mep}
\end{figure}

We design simulated MEP pores with distinct shapes (Fig.~\ref{fig:mep}a) to examine how geometric confinement influences inclusion reconfiguration. The radius of curvature at each end of the pore, $R_1$, is fixed to confine inclusions of radius $R_{inc} = 0.7 R_1$. At the pore waist, we define curvature radii $R_{21}$, $R_{22}$ and $R_{23}$ for MEP1, MEP2 and MEP3, respectively, and use these to tune the antagonistic curvature experienced by the confined inclusion. MEP1, the pillbox design, has straight sidewalls with an infinite radius of curvature at the waist. MEP2 and MEP3 are peanut-shaped, with waist radii $R_{22} = 5R_1$ and $R_{23} = 3R_1$, respectively. Simulated inclusions centered along $z$ and $x$ are placed at varying locations along $y$ to probe the energy landscape along the pore length. Increasing antagonistic curvature at the waist ($y = 0$) increases the distortion of the director as the inclusion is displaced and should raise the barrier to reconfiguration. We examine the influence of pore depth by considering full height ($F$) and half height ($H$) pores, with depths of  $ 2.4R_{inc}$ and $1.2 R_{inc}$, respectively.

Homeotropic inclusions confined within homeotropic pores generate pronounced splay in the surrounding director (Fig.~\ref{fig:mep}b,c). The total elastic free energy as a function of inclusion position along the pore axis (Fig.~\ref{fig:mep}k,l) shows local minima corresponding to stable docking sites for all MEP designs under both full and half height confinement. Displacement of the inclusion along $y$ allows the system to explore two energetically equivalent minima, State A and State B, separated by a barrier that is largest near the pore half-length ($y = 0$), where planar or antagonistic curvature produces the largest elastic penalty. The MEP is thus a bistable mechanical system in which repositioning the inclusion requires an energy input at least an order of magnitude larger than thermal fluctuations, with a barrier height set by pore geometry. For full height confinement, the barriers for MEP1$^F$, MEP2$^F$ and MEP3$^F$ are 11.8, 115.8 and 267.1~$k_BT$, respectively (Fig.~\ref{fig:mep}k). Half height confinement increases the defect area within the pore and yields much higher barriers of 1387.1, 1390.8 and 1452.6~$k_BT$ for MEP1$^H$, MEP2$^H$ and MEP3$^H$ (Fig.~\ref{fig:mep}l). With the three elastic constants of 5CB, full height barriers increase and half height barriers decrease, but every design remains bistable and the ordering with waist curvature is unchanged (Supplementary Table~1). Disclinations in half height MEPs adhere preferentially to sharp edges of the confining boundaries (Fig.~\ref{fig:mep}c), consistent with defect pinning at edges in nematic systems \cite{beller2015}. In the half height case, these defects introduce a discontinuity in the energy landscape at the pore waist, which gives rise to the enhanced barriers.

\subsection*{Experimental MEPs}

\begin{table}[t]
\centering
\caption{Experimental MEP switching behavior. $V_{min}$, minimum voltage at which the inclusion reconfigures in both directions; $E_{min} = V_{min}/d$, where $d$ is the electrode separation measured in each cell (average 17.6~mm); $t_{switch}$, time for the inclusion to travel between docking sites at $E_{min}$. $^*$Uncertainty, $\pm 1\,\mathrm{s}$.}
\begin{tabular}{lccc}
\hline
 & $V_{min}$ (V) & $E_{min}$ (V\,m$^{-1}$) & $t^*_{switch}$ (s) \\
\hline
MEP1$^F$ & 100 & $5.7\times10^3$ & 69 \\
MEP2$^F$ & 250 & $1.4\times10^4$ & 53 \\
MEP3$^F$ & 550 & $3.2\times10^4$ & 24 \\
MEP1$^H$ & 450 & $2.6\times10^4$ & 61 \\
MEP2$^H$ & 450 & $2.5\times10^4$ & 32 \\
MEP3$^H$ & 1200 & $6.9\times10^4$ & 31 \\
\hline
\end{tabular}
\label{tab:switch}
\end{table}

Experimental MEPs are designed analogously to the simulated ones. The radii of curvature at the pore ends (defined by the docking circle) are fixed, and the curvature at the waist is varied systematically (Fig.~\ref{fig:mep}d). Experimental full and half height pores are 10.2 and 5.5~\textmu m deep,  $2.3R_{inc}$ and $1.2R_{inc}$ for inclusions of radius 4.5~\textmu m. Figure~\ref{fig:mep}e--j shows the key configurations of Ag-coated silica spheres of radius $\sim$4.5~\textmu m in full and half height MEPs. In the full height case, the empty pore contains a defect, and the configuration after inclusion assembly is defect-free. In the half height case, empty pores are defect-free, and the docked inclusion appears to be accompanied by an asymmetrically oriented defect. Inclusions entering pores that contain a defect can take its place or repel it, and so serve as probes of the confined defect (Supplementary Fig.~1). Docking is lost when 5CB is heated into the isotropic phase, confirming that it requires nematic order (Supplementary Fig.~2). These configurations are similar to the simulations, with two differences. Firstly, the full height simulations do not include the NLC layer above the pore that is present experimentally, where the copper electrodes at the edges of the cell also serve as spacers between the patterned substrate and the lid (see Methods). Secondly, the half height simulations in the one-constant approximation predict a defect accompanying the inclusion; we hypothesize a distorted line defect encircling the inclusion, akin to the Saturn-ring defect observed around homeotropic microspheres in nematic liquid crystals \cite{gu2000}.

Switching fields are applied in the plane of the cell, along the long axis of the pores, between copper electrodes at the cell edges which are separated by approximately 17.6~mm (Fig.~\ref{fig:concept}c, Methods). Inclusions translate toward the positive electrode and reverse direction when the polarity is reversed, indicating linear electrophoresis of an inclusion carrying a negative effective surface charge, with $F_{electric} \sim qE$. In the absence of an applied field, inclusions sit at the minimum-energy locations predicted by simulation. A DC field translates them to the opposite docking site, and after the field is removed they remain in their new positions, confirming non-volatile bistability (Fig.~\ref{fig:mep}o). In each of the six designs, inclusions can be toggled from State A to State B and back by reversing the field polarity, with similar responses in both directions (Supplementary Fig.~3). The minimum switching fields, $E_{min} = 5.7\times10^3$ to $6.9\times10^4$~V\,m$^{-1}$, follow the trends predicted by simulation, and at these fields inclusions travel between docking sites in $t_{switch} = 24$ to 69~s (Table~\ref{tab:switch}). Below $E_{min}$, inclusions are displaced weakly from their docking sites and then stop. For full height MEPs, the threshold increases from MEP1$^F$ to MEP3$^F$ as antagonistic curvature at the waist increases. For half height MEPs, the thresholds of MEP1$^H$ and MEP2$^H$ are similar, while that of MEP3$^H$ is significantly higher.

Figure~\ref{fig:mep}p shows the dynamics of field-driven reconfiguration for MEP1$^F$ and MEP1$^H$, separated into five regions for which the average inclusion velocity is reported. 

In MEP1$^H$, the velocity rises to 0.62~\textmu m\,s$^{-1}$ when the field is applied (region 1). It then slows  to 0.14 and 0.04~\textmu m\,s$^{-1}$ in regions 2 and 3, respectively, during which the defect accompanying the inclusion undergoes a transition, and subsequently recovers to 0.48~\textmu m\,s$^{-1}$ in region 5.  During the defect transition,  the work done by the field may be consumed by defect restructuring rather than inclusion motion. In MEP1$^F$, which has no accompanying defect, the velocity remains between 0.16 and 0.26~\textmu m\,s$^{-1}$ through regions 1 to 4. In region 5, the inclusions in both MEPs move at approximately the same velocity, indicating that they experience approximately the same elastic force; having passed the pore center, they have overcome the elastic attraction to the original docking site. Inclusion inertia is negligible (see Methods), so the relevant forces are the electric force $F_{electric}$, the elastic force $F_{elastic}$ and the viscous drag $F_{drag}$, which sum to zero,
\begin{equation}
\sum F = F_{electric} + F_{elastic} + F_{drag} \approx 0 .
\end{equation}
Electric fields have been used to translate colloids in nematic liquid crystals \cite{loudet2001,lavrentovich2014}, including by nonlinear electrophoresis, in which the velocity scales with the square of the field and is independent of its polarity \cite{lavrentovich2010}.

To quantify the forces and energy dissipation experienced by inclusions as they relax to equilibrium, we construct experimental elastic energy profiles (Fig.~\ref{fig:mep}m,n). Inclusions are displaced from equilibrium by an applied field, the field is removed, and their spontaneous return to the docking sites is recorded. Migration in MEPs is deterministic, strongly guided by gradients in the director, and Brownian motion can be neglected (see Methods). If inclusions move slowly enough that the NLC equilibrates around them, gradients in the free energy give the forces on the inclusions, and the energy dissipated along a trajectory is
\begin{equation}
E_{elastic} = \int_{y_0}^{y_t} -C_D^{*}\, U \, dy ,
\end{equation}
where $U$ is the inclusion velocity. Inclusions are driven to the unstable position at the pore half-length or waist using the minimum switching voltage, the field is removed, and the displacement $\Delta y$ of the inclusion center of mass is recorded versus time $t$. Velocities, forces and dissipated energies are then computed versus $\Delta y$. Drag coefficients depend on the distance of the inclusion from the pore ends \cite{luo2016}; over the range sampled here these variations are weak, and we take $C_D^{*}$ as a spatially invariant effective drag coefficient,
\begin{equation}
C_D^{*} = C_D\, 6\pi \eta_{\perp} R_{inc} ,
\end{equation}
with $C_D \approx 4$ to account for motion in thin gaps near confining walls and $\eta_{\perp} = 14.3$~mPa\,s, a typical viscosity for motion along the nematogen short axis in 5CB \cite{luo2016}. This gives $C_D^{*} = 4.85\times10^{-6}$~Pa\,s\,m.

Relaxation dynamics depend on MEP geometry. In all cases, inclusions displaced from the unstable position migrate toward the docking sites, accelerating through regions of steep energy gradient and slowing near the minima. Because the forces are proportional to the translational speeds, the inferred forces peak between the waist and the docking site and diminish near the minimum (Supplementary Figs.~4 and 5). MEPs with larger barriers require stronger fields for reconfiguration, leaving the director more distorted at the waist when relaxation begins; hence, although simulations predict vanishing elastic forces at the waist, the inferred force profiles have non-zero slopes there for higher-barrier designs.

Simulations are performed at length scales smaller than those of the experiments. Landau--de Gennes modeling must resolve the nematic correlation length, which sets the size of defect cores and is of order 10~nm, so the mesh spacing is fixed at a fraction of this length \cite{sussman2019}. Resolving pores and inclusions tens of micrometers in size at this spacing is computationally prohibitive. We therefore compare simulations and experiments qualitatively, through the topology of the director configurations and the trends in reconfiguration barriers with pore geometry, rather than through absolute energies. For full height MEPs, the inferred energy profiles reproduce the principal features of the simulations, namely a maximum at the waist and barriers increasing in the order MEP1$^F$, MEP2$^F$, MEP3$^F$. In half height MEPs, defect reconfiguration during inclusion motion produces non-equilibrium effects, non-monotonic forces, and energy profiles without the abrupt features predicted by static simulations, reflecting the finite relaxation times of defects. In experiment, as in the simulations, the barriers of MEP1$^H$ and MEP2$^H$ are similar, while that of MEP3$^H$ is significantly higher. At the scale studied in experiment, all designs exhibit two stable docking sites separated by substantial barriers of order $10^3\,k_BT$, which increase with antagonistic waist curvature.

\subsection*{Static realizations of MEP states}
The elasticity-mediated multistability of MEPs is established above with spherical colloidal inclusions. Because the scattering of a sphere does not depend on its orientation, the optical state of a sphere-based MEP array is set entirely by the inclusion positions. To isolate the optical consequences of MEP multistability from the separate challenge of filling large arrays with colloids, we fabricate static realizations of the two positional states. Deterministic placement of spherical inclusions in large-area pore arrays remains challenging. We deposit lithographically defined Ag chip inclusions, matched in size to the switched colloids, at the two MEP-defined positions within each pore. The chips serve as static scatterers of defined cross-section that reproduce the inclusion arrangements of State A and State B.

\begin{figure}[t]
\centering
\includegraphics[width=0.62\textwidth]{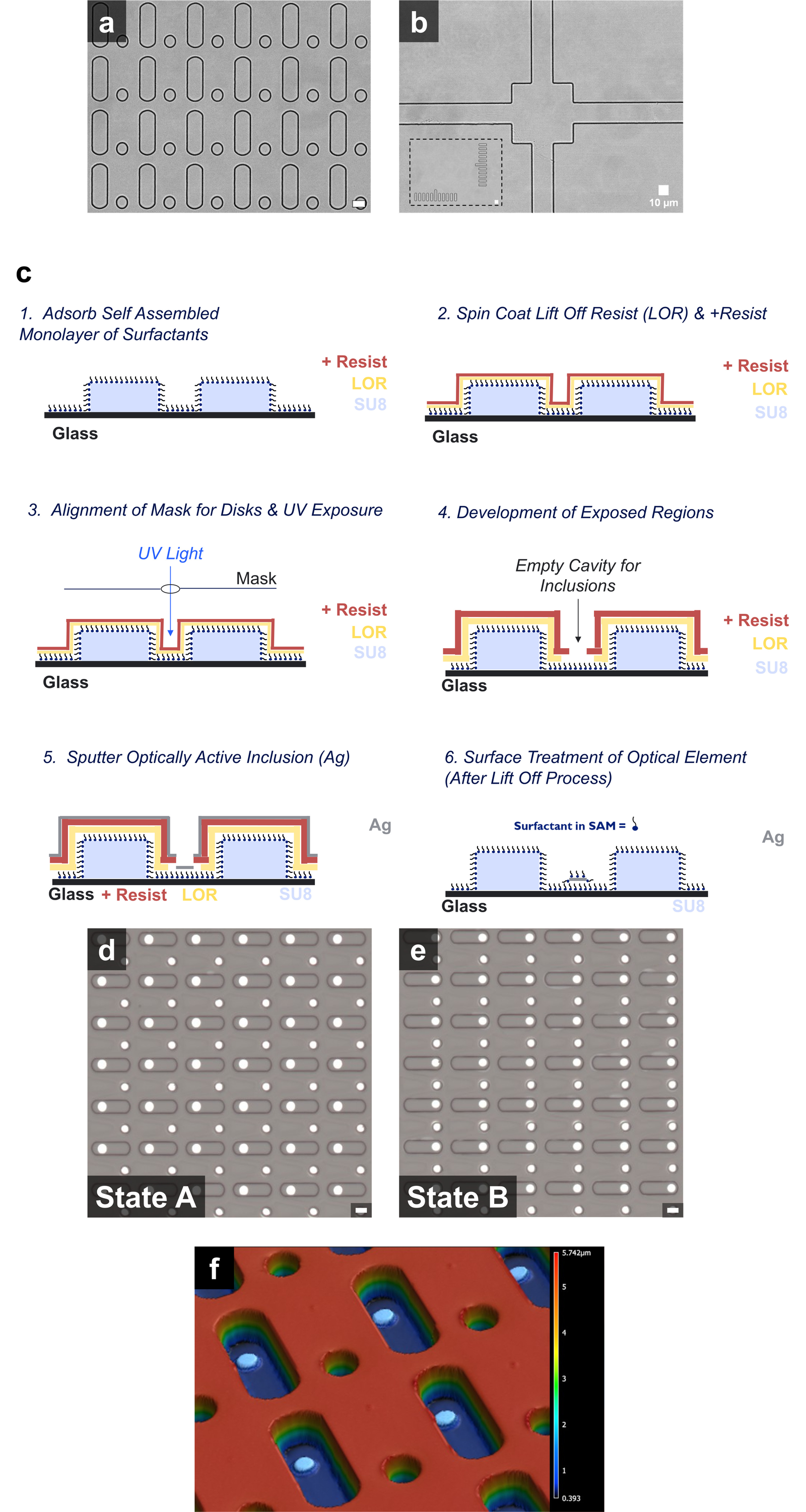}
\caption{\textbf{Static realizations of the two positional states.} \textbf{a} Arrays of MEP and circular pores in SU-8 on glass. \textbf{b} Alignment crosses and verniers. \textbf{c} Fabrication sequence (surfactant SAM, lift-off resist and positive resist, aligned exposure, development, Ag deposition, lift-off and SAM treatment of the chips). \textbf{d,e} Ag chips in MEP and circular pores in the State A (\textbf{d}) and State B (\textbf{e}) configurations. \textbf{f} Laser scanning confocal view of chips in MEP pores. Scale bars, 10~\textmu m. }
\label{fig:fab}
\end{figure}

The fabrication employs multiple mutually aligned photomasks defining arrays of empty MEP and circular pores, MEP pores containing chip inclusions, and circular pores containing chip inclusions (Fig.~\ref{fig:fab}c). The first mask defines arrays of MEP and circular pores in photoresist on glass; these pores form the scaffold of the diffractive array (Fig.~\ref{fig:fab}a). The pore surfaces are then treated with a self-assembled monolayer (SAM) to impose homeotropic anchoring. Alignment crosses and verniers (Fig.~\ref{fig:fab}b) are used to register subsequent masks that define where chips are deposited. Standard photolithography, electron-beam evaporation and lift-off deposit Ag chips $\sim$500~nm thick and of radius $\sim$4.5~\textmu m into selected pores, with separate masks for the two states. A further SAM treatment imposes homeotropic anchoring on the chip surfaces. Figure~\ref{fig:fab}d,e show Ag chips in MEP and circular pores in the State A (lattice) and State B (chain) configurations; the approach places chips with high yield across large-area arrays. Confocal microscopy shows the topography of chips within MEP pores (Fig.~\ref{fig:fab}f).

These static arrays model the two positional states of an MEP diffractive array; they are not themselves reconfigurable. They allow the near-field and far-field signatures of each state to be measured independently of reconfiguration dynamics. In the array-factor description of diffraction, the locations of the diffraction orders are set by the positions of the scatterers alone \cite{balanis2005}. A planar Ag chip and an Ag-coated sphere differ in their single-element responses, which affect the intensities of the orders but not their locations. The static arrays therefore reproduce where the state-dependent diffraction features of a sphere-based MEP array would fall, but not their absolute intensities.

\subsection*{Optical responses of MEP-defined states}

\begin{figure}[t]
\centering
\includegraphics[width=\textwidth]{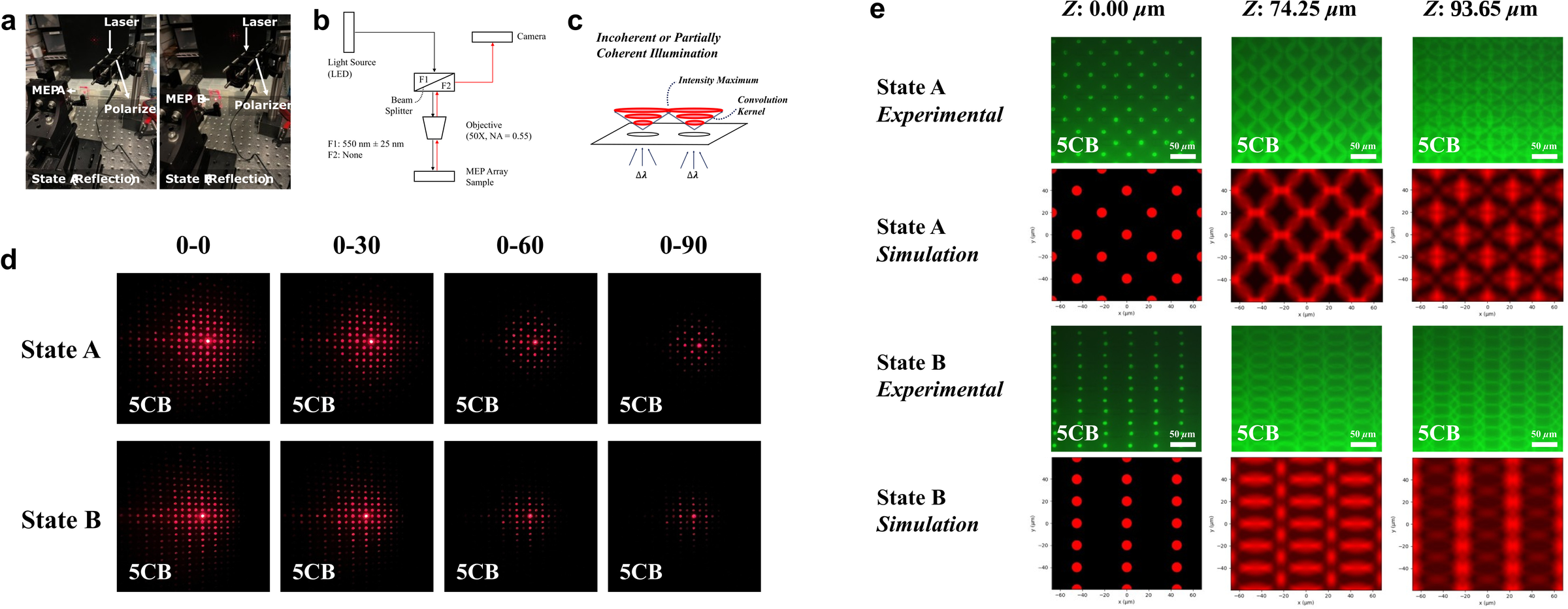}
\caption{\textbf{Optical responses of the two positional states.} \textbf{a} Far-field setups for States A and B in reflection. \textbf{b} Near-field reflection microscopy. \textbf{c} Partially coherent imaging model with a convolution kernel. \textbf{d} Measured far-field patterns of States A and B for polarizer angles of 0, 30, 60 and 90$^\circ$ relative to the incident polarization. \textbf{e} Near-field intensity at three heights $z$, experiment and model, States A and B. }
\label{fig:optics}
\end{figure}

Two NLC cells were assembled from the static arrays shown in Fig.~\ref{fig:fab}d,e, one in the State A configuration and one in State B, using the same anchoring conditions and NLC as in the switching experiments. We define the unit cells corresponding to the two states. In State A, the unit cell consists of a central inclusion with partial contributions from neighboring inclusions at each corner, forming a lattice-like arrangement. In State B, the unit cell contains a central inclusion with two neighboring inclusions above and below, forming a chain-like arrangement. MEPs can in principle operate in the homogenized limit, for inclusion sizes and periodicities much smaller than the wavelength, or in the diffractive limit, for sizes and periodicities comparable to the wavelength. For ease of fabrication, we study diffraction-limited devices, in which the inclusions behave as apertures in reflection. The far-field intensity of a single circular aperture in the Fraunhofer limit follows an Airy profile. For periodic arrangements of inclusions, the far-field pattern is the product of the single-inclusion response and an array factor that depends on the geometry and spacing of the arrangement \cite{balanis2005}. We use this framework to guide our interpretation of the far-field responses.

We measured the far-field responses of the two states in reflection using a laser ($\lambda \sim 630$~nm) obliquely incident on the cells (Fig.~\ref{fig:optics}a). A linear polarizer in the detection path, oriented at various angles relative to the incident linear polarization, reduces background scattering from the pores; the highest contrast is obtained with the polarizer at 90$^\circ$. The measured patterns (Fig.~\ref{fig:optics}d) correspond to the periodicities of the respective unit cells, consistent with Fraunhofer and array-factor descriptions, with oblique incidence shifting the diffraction orders laterally.

In transmission, the far-field patterns of the two states are nearly indistinguishable. At the diffractive scale explored, light transmitted through the transparent cell is diffracted more strongly by the pore lattice, which is common to both states, than by the inclusions. In reflection, the angular spread of the diffracted light is instead set by the effective aperture of each reflective inclusion, and the state-dependent orders are resolved. All optical measurements reported here were therefore performed in reflection.

In the near field, the two states also give distinct responses (Fig.~\ref{fig:optics}e). Samples were illuminated from above with an LED filtered to $550 \pm 25$~nm, and the reflected light was collected (Fig.~\ref{fig:optics}b) with an objective of numerical aperture NA$_{obs} = 0.55$. In microscopy the illumination NA is generally smaller than the observation NA \cite{maurer2008}; we take NA$_{ill} = 0.28$. The bandpass filter gives a coherence length $l_c = \lambda_0^2/\Delta\lambda \approx 6.05$~\textmu m and coherence time $\tau_c = l_c/c \approx 20$~fs. The LED emits from many uncorrelated points, giving low spatial coherence and a broad angular distribution of incident waves; the illumination is therefore partially coherent, and the Fresnel pattern expected for coherent plane-wave illumination is not observed \cite{goodman2005}. Instead, intensity is concentrated between adjacent inclusions, consistent with the intersection of circularly blurred replicas of the structure expected for incoherent imaging. We therefore model each inclusion as independently illuminated and add intensities rather than fields. In each plane $z$, the structure is convolved with a circular kernel of radius $r = z\,\mathrm{NA}_{ill}$, the intensity point spread function of the system, and local maxima appear where kernels intersect (Fig.~\ref{fig:optics}c; see Methods). This empirical model reproduces the measured near-field intensity distributions, which differ markedly between States A and B. Thus, the far-field and near-field measurements show that the two MEP-defined positional states correspond to distinct optical responses at visible wavelengths.

\section*{Discussion}

Our simulations and experiments establish that elastic energy landscapes in confined NLCs can be engineered to store the position of a scattering inclusion in mechanically stable, non-volatile configurations, and that the resulting positional states correspond to distinct optical responses. The MEP mechanism differs from established routes to reconfiguration. In LC-tuned metasurfaces, the director surrounding fixed resonators is reoriented and relaxes when the field is removed \cite{li2019,liJ2020}. In phase-change metasurfaces, the state is stored in the metastable phase of a film adjacent to fixed resonators, and switching requires melting or annealing \cite{fang2024}. In MEPs, the optical element itself is relocated by a transient field and held by an elastic barrier. No material undergoes a phase transition, and because reconfigurability is governed by geometry-defined elastic landscapes, the composition of the inclusion can be chosen for optical function alone.

These features suit MEPs to applications in which an optical configuration must be held for long periods and changed only occasionally, such as reconfigurable diffractive elements, configurable filters and gratings, and non-volatile optical memory. MEP states are held by elastic barriers far exceeding thermal energies and require no energy input between switching events, the regime exploited by bistable nematic displays to retain images without power \cite{zbd2022}. Because the switching threshold is set by pore geometry, MEPs also offer a route to programmable devices. In full height MEPs, the threshold field increases more than fivefold from MEP1$^F$ to MEP3$^F$ as the antagonistic curvature at the waist increases (Table~\ref{tab:switch}), so the threshold responds robustly to changes in MEP shape. In an array containing several MEP designs, a field above all thresholds sets every pore to one state, and a reversed field between two thresholds then moves only the lower-threshold pores, writing distinct subsets of the array without patterned electrodes. Random partial filling adds a further level of encoding, fixed at fabrication and difficult to reproduce, while the inclusion positions in occupied pores remain rewritable. Threshold selection and partial filling can yield many distinct optical signatures within a single device, suited to tagging and anti-counterfeiting. Where pixel-level control is needed, electrodes patterned at the scale of each pore, as demonstrated for electrochemically switched metasurfaces \cite{kaissner2024}, would allow individual unit cells to be written. Because each inclusion switches only above a threshold field, fringing fields at neighboring pores should leave them in their stored states, and non-volatile pixels with threshold switching can be written row by row in a passive-matrix architecture without the transistor backplane required for volatile pixels \cite{gu2023}. While here we have focused on electric switching fields, magnetic switching fields can be used to switch ferromagnetic colloids between bistable MEP states (Supplementary Fig.~6). 

MEPs can also be scaled into the metasurface regime. In the strong anchoring regime, the reconfiguration barrier scales linearly with inclusion radius, $\Delta E = c\,KR$, where the prefactor $c$ is set by pore geometry and is of order unity for peanut-shaped MEPs. The simulations themselves operate at the nanoscale. With inclusions of radius 94.5~nm, full height MEPs present barriers of 12 to 267~$k_BT$ (Fig.~\ref{fig:mep}k). With an attempt time of order $\eta R^3/k_BT \sim 10^{-2}$~s, barriers above roughly 30~$k_BT$ retain a state for years against thermal escape, so peanut-shaped MEPs remain bistable at this scale and waist curvature becomes the key design parameter. Arranged on a subwavelength lattice, MEPs of these dimensions would bring the approach from the diffractive regime explored here into the metasurface regime at near-infrared wavelengths, in which each pore acts as a meta-atom whose optical response is set by the position of its inclusion. Elastic coupling between inclusions and the nematic director has been demonstrated for particles of radius $\sim$60~nm \cite{skarabot2010}, comparable to the simulated inclusions, and field-driven transport of particles in nematic liquid crystals is well established \cite{lavrentovich2014}, so both docking and electrophoretic switching should remain available as MEPs are reduced in size. The MEP concept is limited only as the inclusion approaches the nematic coherence length, of order 10~nm \cite{degennes1995}, where the order parameter is reduced rather than the director distorted, or as the pore shrinks to the size at which distinct stable states give way to a single order-reconstructed state \cite{kralj2014}.

Several directions follow directly from this work. Switching of filled MEP arrays with simultaneous optical measurement awaits methods for deterministic filling of large pore arrays with colloids in an NLC; the static arrays reported here establish the optical contrast that such arrays will deliver. The state-independent background of the pore lattice can be reduced by matching the refractive index of the pore material to the ordinary index of the NLC, by reducing the pore period below the wavelength, or by designing inclusion arrangements whose diffraction orders are absent from the pore lattice. Anisotropic inclusions and complex confining pores offer additional degrees of freedom; designing shape, anchoring and pore geometry together could enable, for example, chiral or polarization-selective responses.

MEPs establish nematic elasticity as a means of storing the position of an optical element, rather than the state of an optical medium. This positional mode of non-volatile reconfiguration complements phase-change and LC-tuned approaches, and it is compatible in principle with scatterers of arbitrary composition at any wavelength for which the NLC is transparent.

\section*{Methods}

\subsection*{Simulations}

We simulate the relaxation of nematic liquid crystals confined within MEP pores using a Landau--de Gennes Q-tensor (LdG) framework \cite{ravnik2009,beller2014thesis}. All simulations were performed with the original Qmin code. The related open-source code open-Qmin, which shares its finite-difference foundation, is publicly available \cite{sussman2019}. The total free energy is minimized on a uniform cubic mesh by a finite-difference method, using the conjugate gradient algorithm of the ALGLIB library (www.alglib.net). The LdG free energy functional comprises bulk nematic phase ($f_{\mathrm{phase}}$) and elastic distortion ($f_{\mathrm{d}}$) contributions, where $L_1$, $L_2$, and $L_3$ are the LdG elastic coefficients expressed in terms of the Frank elastic constants $K_1$, $K_2$, and $K_3$.

\begin{align}
F_{\mathrm{LdG}} &= \int_V dV \left[f_{\mathrm{phase}} + f_{\mathrm{d}}\right] \\
f_{\mathrm{phase}} &= \frac{1}{2}A\operatorname{tr}(\mathbf{Q}^2)
 + \frac{1}{3}B\operatorname{tr}(\mathbf{Q}^3)
 + \frac{1}{4}C\left(\operatorname{tr}(\mathbf{Q}^2)\right)^2 \\
f_{\mathrm{d}} &= \frac{1}{2}L_1 \frac{\partial Q_{ij}}{\partial x_k}\frac{\partial Q_{ij}}{\partial x_k}
 + \frac{1}{2}L_2 \frac{\partial Q_{ij}}{\partial x_j}\frac{\partial Q_{ik}}{\partial x_k}
 + \frac{1}{2}L_3 Q_{ij}\frac{\partial Q_{kl}}{\partial x_i}\frac{\partial Q_{kl}}{\partial x_j} \\
L_1 &= \frac{(2/3)(-K_1 + 3K_2 + K_3)}{9S^2}, \quad
L_2 = \frac{4(K_1-K_2)}{9S^2}, \quad
L_3 = \frac{4(K_3-K_1)}{27S^3}
\end{align}

The bulk LdG free-energy coefficients for 5CB were set to $A = -0.172 \times 10^6\,\mathrm{J\,m^{-3}}$, $B = -2.12 \times 10^6\,\mathrm{J\,m^{-3}}$, and $C = 1.73 \times 10^6\,\mathrm{J\,m^{-3}}$, corresponding to an equilibrium scalar order parameter of $S_0 \approx 0.533$. Strong homeotropic anchoring was imposed with an anchoring strength of $W_0 \approx 1 \times 10^{-2}\,\mathrm{J\,m^{-2}}$. Simulated pores consist of two docking circles of radius $R_1$ whose centers are separated by $2R_1$, giving a pore of length $4R_1$ and width $2R_1$, with waist radii $R_{22} = 5R_1$ (MEP2) and $R_{23} = 3R_1$ (MEP3). Inclusions have radius $R_{inc} = 0.7R_1 = 94.5$~nm, so that one mesh spacing corresponds to 4.5~nm. The computational domain spans $100 \times 130 \times 50$ mesh spacings in $x$, $y$ and $z$, with $R_1 = 30$ and $R_{inc} = 21$ spacings. Full-height pore walls span the domain thickness (50 spacings, $2.4R_{inc}$), and half-height pore walls are 25 spacings ($1.2R_{inc}$) deep, close to the experimental ratios of $2.3R_{inc}$ and $1.2R_{inc}$. All surfaces impose strong homeotropic anchoring. The inclusion is displaced along $y$ in steps of one mesh spacing, and the free energy is minimized at each position. In the one-constant approximation, $K = K_1 = K_2 = K_3 = 1\times10^{-11}$~N. In the three-constant approximation, $K_1 = 0.64\times10^{-11}$~N, $K_2 = 0.3\times10^{-11}$~N and $K_3 = 1\times10^{-11}$~N, the values for 5CB. In both, the saddle-splay constant is $K_{24} = K_2$. Pore boundaries are defined as boundary files in MATLAB and imported into the code, and the coordinates of MEP1, MEP2 and MEP3 are computed in Mathematica (Supplementary Information).

\subsection*{Electric field-driven motion}
Masks defining the pores for filled MEP experiments were designed in LayoutEditor and fabricated by direct-write lithography. Pores were fabricated on 3- or 4-inch Borofloat glass wafers in negative-tone epoxy photoresist (SU8 3005 and SU8 3010, Kayaku Advanced Materials) by standard lithography. Experimental pores have docking circles of radius $R_1$ with $R_{inc}/R_1 = 0.6$ and $L/R_1 = 3.7$, where $L$ is the center-to-center distance between the docking circles. The docking radius $R_1= 7.2$~\textmu m, giving an overall pore length $L + 2R_1  \approx 41$~\textmu m; waist radii $R_{22}/R_1 = 11.6$ (MEP2) and $R_{23}/R_1 = 7.4$ (MEP3) (Fig.~\ref{fig:mep}d). Average pore depths for full and half height MEPs were $10.21 \pm 0.05$~\textmu m and $5.49 \pm 0.23$~\textmu m, respectively. Substrates were cleaned with O$_2$ plasma, and a SiO$_2$ layer was deposited by vapor deposition from SiCl$_4$ or by sputtering ($\sim$20~nm). NLC cells were assembled with 25~\textmu m copper spacers, giving cell thicknesses of approximately 32 to 34~\textmu m. The copper spacers, placed at opposite edges of the cell and separated by approximately 17.6~mm, serve as electrodes, and pores are oriented with their long axis parallel to the applied field. Top glass slides were cleaned with 2\% aqueous Micro-90, rinsed with deionized (DI) water, immersed in Piranha solution (30\% aqueous H$_2$O$_2$ and concentrated H$_2$SO$_4$, 1:3 v/v) at 80~$^\circ$C for 1~h, rinsed, dried and stored under low humidity. Colloidal inclusions were Ag-coated monodisperse silica microspheres of radius 4.5~\textmu m (nominal diameter 9~\textmu m, Cospheric), cleaned in 1:2 v/v IPA and DI water, treated to impose homeotropic anchoring, and suspended in 5CB (Nematel). All surfaces (pores, glass, inclusions) were treated with a 3~wt\% aqueous solution of dimethyloctadecyl[3-(trimethoxysilyl)propyl]ammonium chloride (DMOAP). Colloids were immersed in DMOAP at 80~$^\circ$C for 1~h; pores and lids were immersed for 10~min, rinsed with DI water and baked at 80~$^\circ$C for 1~h. Cells were filled with the colloid suspension in the isotropic phase ($\sim$50 to 60~$^\circ$C) and studied by upright microscopy (Zeiss). The electrodes were connected to an external DC voltage source.

\subsection*{NLC director field-driven motion}
Migration experiments were conducted at or near the lowest switching voltage, defined as the voltage at which the inclusion reconfigures in both directions. This voltage was used to translate the inclusion to the pore half-length or waist, after which the field was removed and the migration recorded. Images were inverted and thresholded in Fiji, and the inclusion center of mass along $y$ was tracked with TrackMate~7 \cite{ershov2022} (linking and gap-closing maximum distances 15~\textmu m; maximum frame gap 0). Trajectories were fitted with the MATLAB Curve Fitting Toolbox. Full height trajectories were well described by the Gompertz sigmoidal model,
\begin{equation}
y = f(t) = d + (a - d)\, e^{-e^{-b(t-c)}} ,
\end{equation}
and half height trajectories by a smoothing spline,
\begin{equation}
p \sum_i w_i \left(y_i - s(x_i)\right)^2 + (1-p) \int \left(\frac{d^2 s}{dx^2}\right)^2 dx .
\end{equation}
Fitted trajectories were differentiated numerically to obtain velocities; velocity profiles (multiplied by $-1$) versus displacement were fitted with smoothing splines and integrated to give experimental energy profiles. 

To substantiate the assumptions of negligible inertia and small departures from the equilibrium director field, we take an average inclusion velocity $v = 0.45$~\textmu m\,s$^{-1}$, NLC density $\rho \approx 1.0$~g\,cm$^{-3}$, effective viscosity $\eta = 30$~mPa\,s, $K = 10^{-11}$~N and inclusion radius $R = 4.5$~\textmu m \cite{sengupta2013ijms}. The Reynolds number $Re = \rho v R/\eta \approx 7\times10^{-8}$ places inclusion motion in the creeping-flow regime. The Ericksen number $Er = \eta v R/K = 0.01$ indicates that viscous stresses do not significantly distort the director from equilibrium as the inclusion translates in the absence of a field. Following estimates for micrometer-sized particles in NLCs \cite{lavrentovich2014}, the director relaxation time around the sphere is $\tau = \beta^2 \eta_{\parallel,\perp} R^2/K \approx 0.2$~s, with $\beta$ of order 1 and $\eta_{\parallel,\perp}/K \sim 10^{10}$~s\,m$^{-2}$. For 5CB, $\eta_\perp$ is approximately 1.8 times $\eta_\parallel$. The choice of $\eta_\perp$ follows from the simulated director fields, in which homeotropic anchoring at the lids aligns the NLC predominantly vertically, more so in half height MEPs, so that inclusions translate through vertically aligned nematogens.

\subsection*{Static arrays}
Photomasks for the pore arrays and chip inclusions were designed in LayoutEditor and written onto a 5-inch AZ1500 photomask on a soda lime substrate using a Heidelberg DWL 66+ Laser Writer. Mask 1 defined the MEP and circular pores in negative-tone photoresist (SU8 3005, Kayaku Advanced Materials) on glass. Mask 2 defined chip positions in MEP pores, with separate versions for State A and State B, and Mask 3 defined chip positions in circular pores. After pore fabrication, substrates were cleaned with O$_2$ plasma and sputtered with SiO$_2$ ($\sim$20~nm). Lift-off resist and positive-tone photoresist (S1818 or KPRO3) were spin-coated onto the pore arrays, and Masks 2 and 3 were aligned to the pores using alignment crosses and verniers. Ag chips $\sim$500~nm thick were deposited by evaporation into the exposed openings, followed by lift-off. The samples were then treated with 3~wt\% aqueous DMOAP to impose homeotropic anchoring.

\subsection*{Optics}
Static arrays were assembled into NLC cells with 8~\textmu m Cu spacers and DMOAP-treated (3~wt\%) glass lids and filled with 5CB (Sigma-Aldrich). Near-field measurements used an upright microscope (Zeiss) in reflection with an LED filtered to $550 \pm 25$~nm; the reflected power distributions were recorded and modeled in Python with an empirically determined convolution kernel (Supplementary Information). Far-field measurements used a $\sim$630~nm linearly polarized source and a linear polarizer in the detection path.

\section*{Data availability}
All data supporting the findings of this study are included in the article and its Supplementary Information.
\section*{Code availability}
The Qmin code used for the Landau--de Gennes simulations is available from the corresponding authors upon request. The related open-source code open-Qmin is publicly available \cite{sussman2019}. 

\section*{Acknowledgements}
This work was supported by the Army Research Office under grant W911NF-22-1-0205; J.-J.S.E. was supported by the National Science Foundation Graduate Research Fellowship Program under grant DGE-1845298. This work was carried out in part at the Singh Center for Nanotechnology at the University of Pennsylvania, which is supported by the NSF National Nanotechnology Coordinated Infrastructure Program under grant NNCI-2025608. 
\section*{Author contributions}
J.-J.S.E.  performed  experiments and simulations, and analyzed the data. B.E. guided the diffraction studies. N.E. and K.J.S. directed the research. All authors contributed to writing the manuscript.
\section*{Competing interests}
J.-J.S.E., N.E. and K.J.S. are inventors on US Patent 12,619,003, assigned to the University of Pennsylvania, covering multistable elastic pixel-based reconfigurable metasurfaces. B.E. declares no competing interests.

\bibliographystyle{unsrtnat}
\bibliography{refs}

\clearpage
\section*{Supplementary Information}
\setcounter{figure}{0}
\setcounter{table}{0}
\renewcommand{\figurename}{Supplementary Fig.}
\renewcommand{\tablename}{Supplementary Table}
\def\theHfigure{S\arabic{figure}}
\def\theHtable{S\arabic{table}}
\newcounter{snote}
\newcommand{\snote}[1]{\refstepcounter{snote}\subsection*{Supplementary Note~\thesnote: #1}}

\snote{one- and three-constant simulations}\label{note:onethree}
Simulations were repeated in the three-constant approximation using the elastic constants of 5CB ($K_1 = 0.64$, $K_2 = 0.3$ and $K_3 = 1$, in units of $10^{-11}$~N; $K_{24} = K_2$). In full height MEPs, the barriers are higher than in the one-constant approximation, owing to the higher cost of bend deformation in 5CB, with the largest relative increase for the pillbox design. In half height MEPs, the barriers are lower, because the inclusion at the waist generates predominantly splay distortion, which is less costly in 5CB. The defect accompanying the inclusion in half height MEPs is pinned at the top edges of the pore in the one-constant approximation and at the bottom edges in the three-constant approximation. In both approximations, every design presents two stable docking sites, and the barriers follow the same ordering with waist curvature (Supplementary Table~\ref{tab:barriers13c}).

\begin{table}[h]
\centering
\caption{\textbf{Simulated reconfiguration barriers $\Delta E_{max}$ ($k_BT$) in the one-constant (1C) and three-constant (3C) approximations.}}
\label{tab:barriers13c}
\begin{tabular}{lcc}
\hline
 & 1C & 3C \\
\hline
MEP1$^F$ & 11.8 & 44.0 \\
MEP2$^F$ & 115.8 & 171.3 \\
MEP3$^F$ & 267.1 & 337.4 \\
MEP1$^H$ & 1387.1 & 1106.7 \\
MEP2$^H$ & 1390.8 & 1147.4 \\
MEP3$^H$ & 1452.6 & 1226.6 \\
\hline
\end{tabular}
\end{table}

\begin{figure}[h]
\centering
\includegraphics[width=0.85\textwidth]{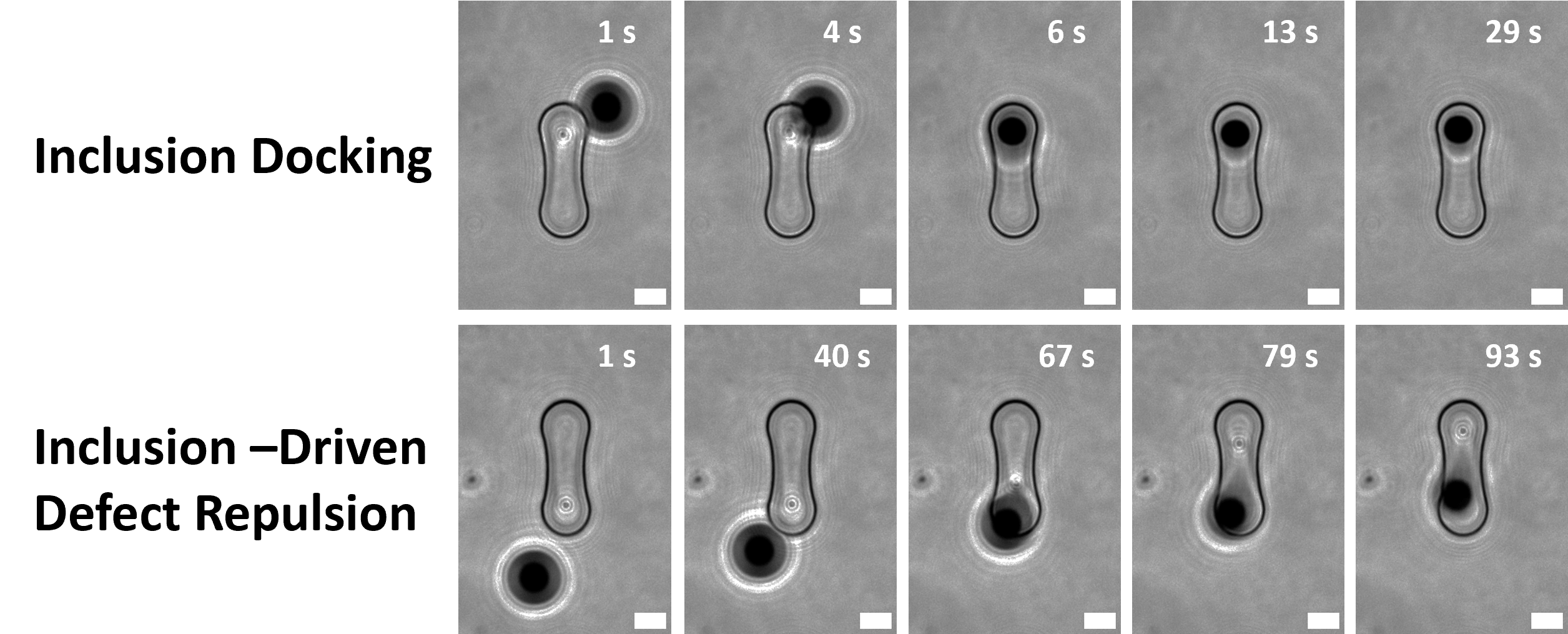}
\caption{\textbf{Inclusions as probes of confined defects.} Top, an inclusion docking into MEP3$^F$ from the NLC cell gap and replacing a point defect. Bottom, an inclusion repelling a point defect in MEP3$^F$. Scale bars, 10~\textmu m.}
\label{fig:particledefectreplacement}
\end{figure}

\snote{Colloidal inclusions as probes of confined defects}\label{note:defectprobes}
In MEP3$^F$, an inclusion with a companion Saturn ring docked from the cell gap into a pore containing a point defect and took its place, indicating that the defect was a radial hedgehog; in another pore, an inclusion repelled a point defect, which may indicate a hyperbolic hedgehog, although the stability of hyperbolic hedgehogs at MEP docking sites has not been verified numerically (Supplementary Fig.~\ref{fig:particledefectreplacement}). An inclusion with specified anchoring can therefore be used to characterize a defect confined within a pore.

\begin{figure}[h]
\centering
\includegraphics[width=0.75\textwidth]{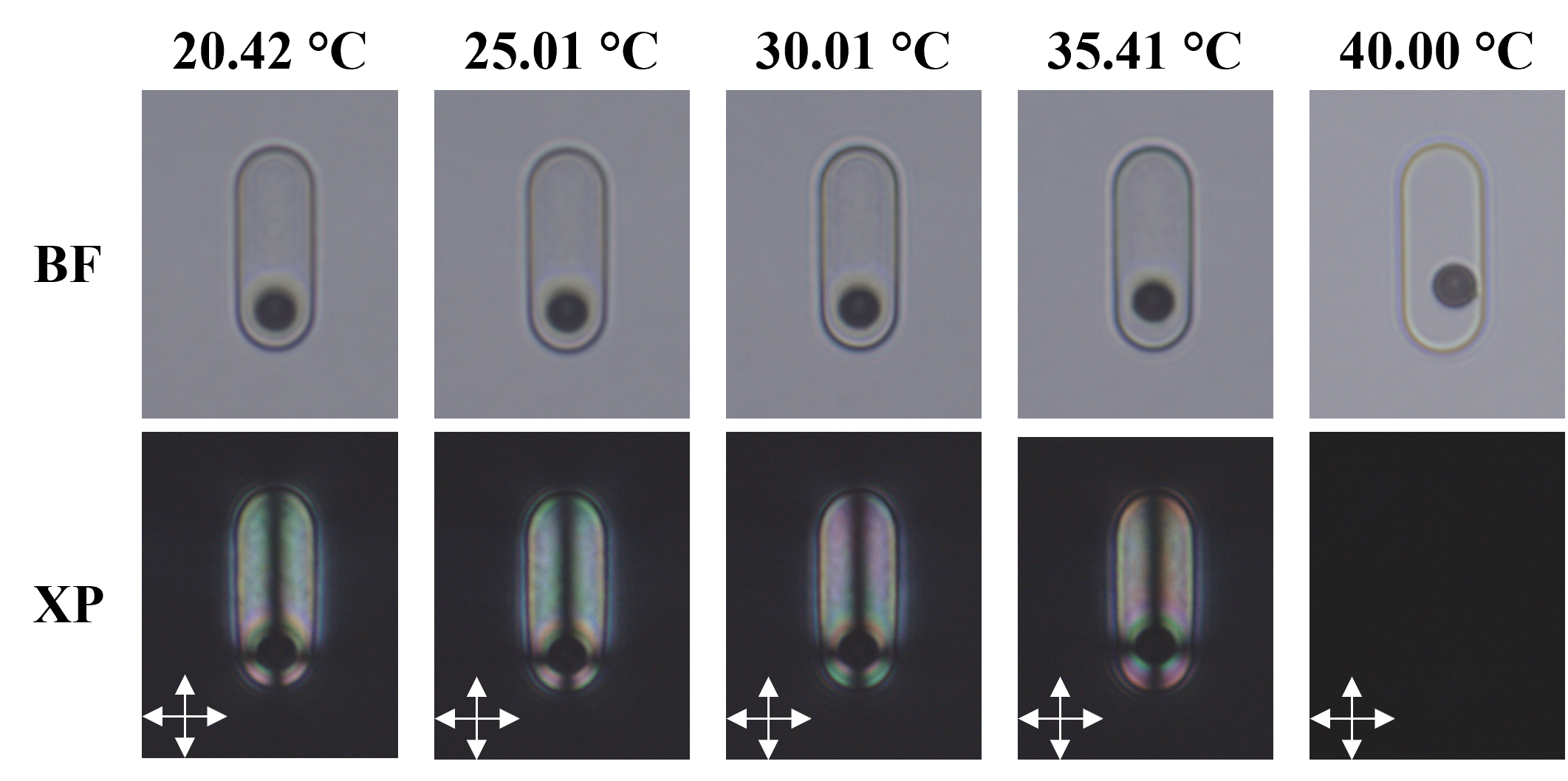}
\caption{\textbf{Temperature-driven release of an inclusion from its docking site in MEP1$^F$.} Bright-field (BF) and crossed-polarizer (XP) images from 20 to 40~$^\circ$C. At 40~$^\circ$C, 5CB is isotropic and docking is lost. Inclusion radius $\sim$4.5~\textmu m.}
\label{meltlosedock}
\end{figure}

\snote{loss of docking in the isotropic phase}\label{note:isotropic}
At 40~$^\circ$C, where 5CB is isotropic, inclusion docking is lost (Supplementary Fig.~\ref{meltlosedock}, MEP1$^F$), confirming that nematic elasticity is required for controlled reconfiguration.

\begin{figure}[h]
\centering
\includegraphics[width=\textwidth]{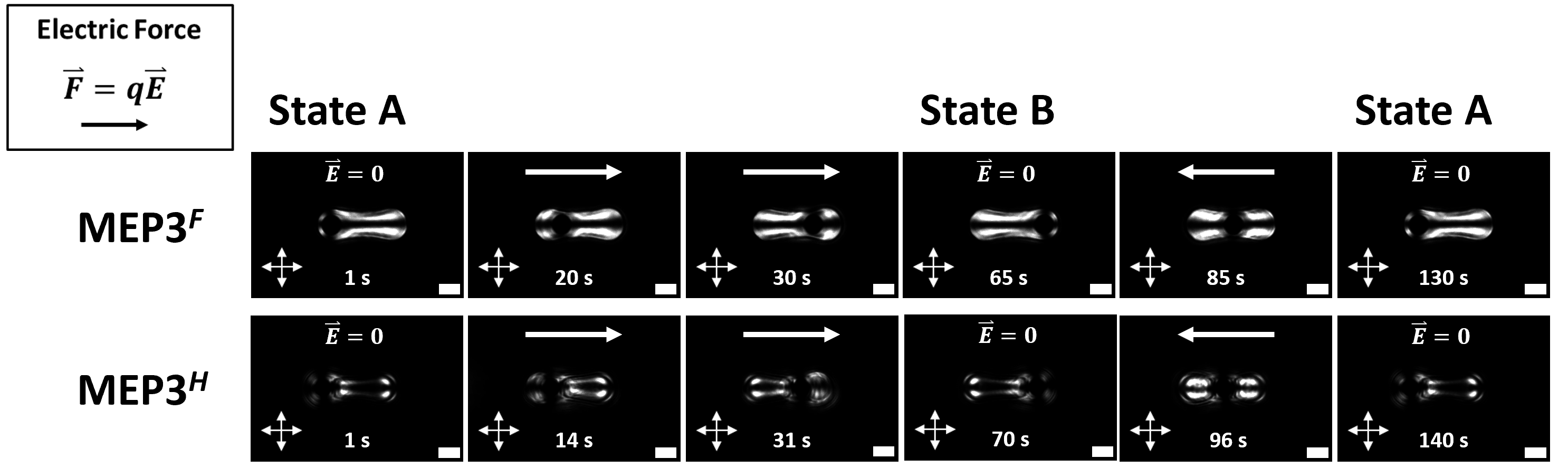}
\caption{\textbf{Crossed-polarizer time series of field-driven reconfiguration} in MEP3$^F$ and MEP3$^H$, State A $\to$ B $\to$ A. Arrows indicate the direction of the applied field.}
\label{fig:xpoltime}
\end{figure}

\snote{director reorganization during switching}\label{note:xpol}
Crossed-polarizer images of field-driven reconfiguration in MEP3$^F$ and MEP3$^H$ at the lowest switching field (Supplementary Fig.~\ref{fig:xpoltime}) show that in full height pores the inclusion reconfigures without significant change in the crossed-polarizer pattern, whereas in half height pores the nematogens reorganize clearly, particularly when the inclusion sits at the waist ($t = 96$~s). Reconfiguration does not lead to irreversible changes in the organization of the pore.

\snote{relaxation dynamics}\label{note:relax}
Trajectories, speeds, and the inferred elastic forces and energies for inclusions relaxing from the waist to a docking site are shown for full height MEPs in Supplementary Fig.~\ref{fig:relaxfull} and for half height MEPs in Supplementary Fig.~\ref{fig:relaxhalf}. Because MEP geometries are symmetric, profiles are taken as identical for inclusions docking at either site. Slight differences are observed in the migration to the left and right docking positions (Supplementary Fig.~\ref{fig:relaxfull}c): the stable location on the left is displaced 6.3~\textmu m from the pore edge and that on the right 6.6~\textmu m, differences that likely originate from slight variations in inclusion size, surface anchoring and pore geometry. In half height MEPs, the inclusion and its accompanying defect pass from a symmetric configuration at the waist ($D_2$) to an asymmetric configuration at the docking site ($D_1$); the transition ($D_{21}$) coincides with a local minimum in speed.

\begin{figure}[p]
\centering
\includegraphics[width=\textwidth]{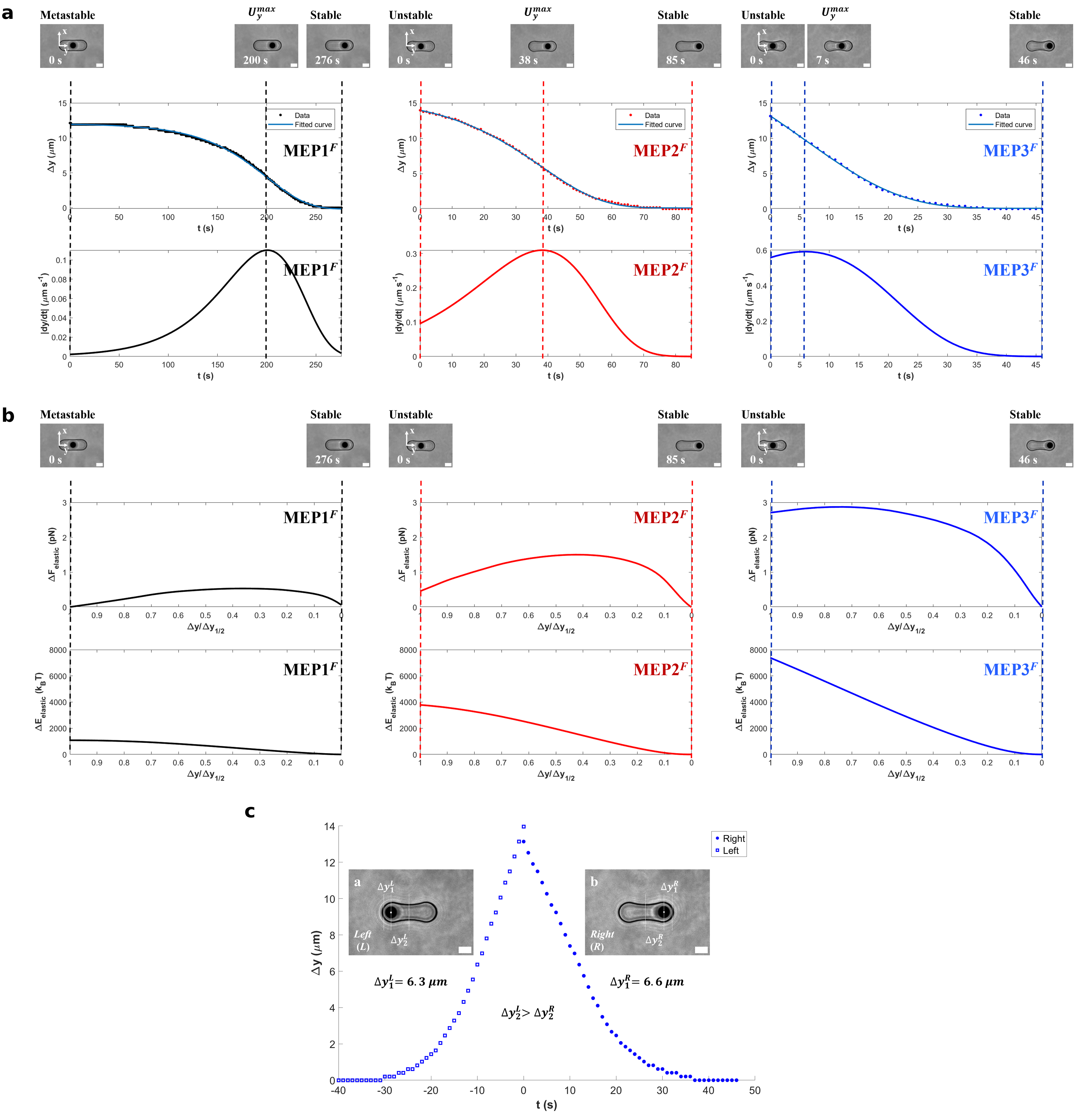}
\caption{\textbf{Relaxation of inclusions in full height MEPs.} Inclusions are driven to the waist at the minimum switching field, the field is removed, and the inclusion relaxes to a docking site. \textbf{a} Displacement $\Delta y$ versus time (points, data; lines, Gompertz fits) and speed $|dy/dt|$ versus time for MEP1$^F$, MEP2$^F$ and MEP3$^F$. Micrographs mark the initial position, the maximum speed $U_y^{max}$ and the final docked state. \textbf{b} Inferred elastic force and elastic energy versus normalized position $\Delta y/\Delta y_{1/2}$. \textbf{c} Relaxation of MEP3$^F$ to the left (open) and right (filled) docking sites. Scale bars, 10~\textmu m.}
\label{fig:relaxfull}
\end{figure}

\begin{figure}[p]
\centering
\includegraphics[width=\textwidth]{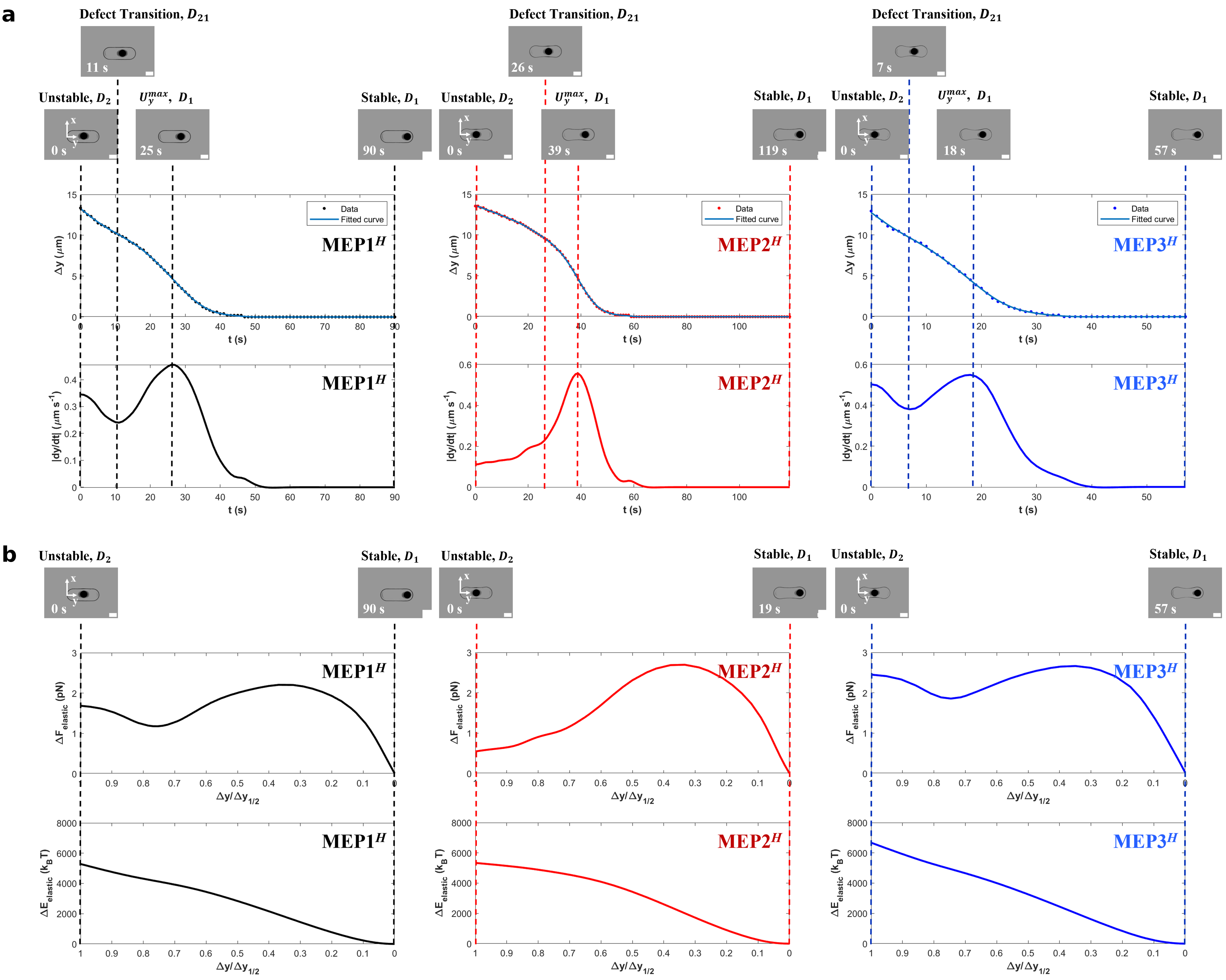}
\caption{\textbf{Relaxation of inclusions in half height MEPs.} As in Supplementary Fig.~\ref{fig:relaxfull}a,b, for MEP1$^H$, MEP2$^H$ and MEP3$^H$ (lines, smoothing-spline fits). Micrographs mark the inclusion and defect configurations. $D_2$, symmetric configuration at the waist; $D_1$, asymmetric configuration at the docking site; $D_{21}$, transition between them, at a local minimum in speed. Scale bars, 10~\textmu m.}
\label{fig:relaxhalf}
\end{figure}

\snote{magnetic switching field}\label{note:magnetic}
MEP reconfiguration utilizing a magnetic switching field was demonstrated using ferromagnetic CrO$_2$-coated polystyrene (CrO$_2$-PS) inclusions of diameter 8.74~\textmu m (SPHERO\texttrademark). After washing with isopropyl alcohol and deionized water, inclusion surfaces were treated with DMOAP (ACROS Organics) to impose homeotropic anchoring. Inclusions were suspended in 5CB (Kingston Chemicals Limited). Standard UV photolithography procedures were performed to fabricate this MEP atop a borosilicate glass wafer (Pure Wafer), and a glass slide was used as the lid for the NLC cell. All surfaces were treated to impose homeotropic anchoring. Strips of Mylar sheets of thickness $\sim$15~\textmu m were utilized as spacers in the NLC cell, resulting in cell thicknesses of $\sim$25--30~\textmu m. The magnetic field, supplied locally by a handheld magnetic probe, provided sufficient field strengths to control inclusion locations. In the presence of the switching field, the inclusion could be moved between docking sites within the MEP. Inclusion positions were stable in either state (A or B) absent the magnetic switching field.

\begin{figure}[p]
\centering
\includegraphics[width=\textwidth]{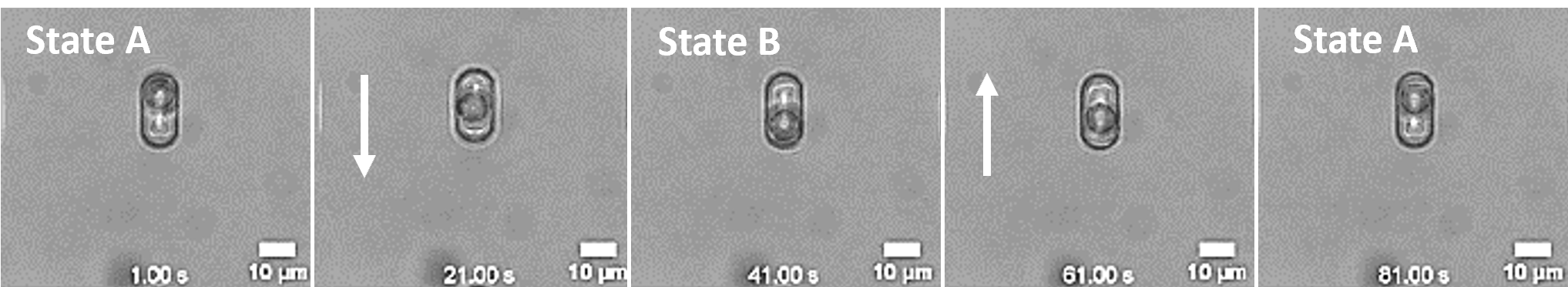}
\caption{\textbf{Magnetic field reconfiguration of ferromagnetic inclusion in pillbox MEP.} Reversible switching, State A $\to$ B $\to$ A of a ferromagnetic inclusion under a transient, local magnetic field. Arrows are a visual aid to illustrate translation direction. Scale bars, 10~\textmu m.}
\label{fig:magnetic}
\end{figure}

\end{document}